# Automated decision-making and artificial intelligence at European borders and their risks for human rights

**Yiran Yang, Frederik Zuiderveen Borgesius, Pascal Beckers, Evelien Brouwer**

*Working draft V. 1. We would love to hear your comments or suggestions. Please contact us before citing. Thank you!*

*Frederikzb@cs.ru.nl*

Author note

Yiran Yang,[a,b] Frederik Zuiderveen Borgesius,[b,c] Pascal Beckers,[a] Evelien Brouwer[d]

[a]Institute for Management Research, Department of Geography, Planning and Environment, Radboud University, Heyendaalseweg 142, 6525 AJ, Nijmegen, The Netherlands

[b]Interdisciplinary hub for Digitalization & Society (iHub), Radboud University, Erasmuslaan 1, 6525 GE, Nijmegen, NL, the Netherlands

[c]Institute for Computing and Information Sciences (iCIS), Radboud University, Toernooiveld 212, 6525 EC Nijmegen, the Netherlands

[d]Institute of Constitutional, Administrative Law and Legal Theory, Utrecht University School of Law, Utrecht University, Newtonlaan 201, 3584 BH, Utrecht, The Netherlands

**Abstract**

Many countries use automated decision-making (ADM) systems, often based on artificial intelligence (AI), to manage migration at their borders. This interdisciplinary paper explores two questions. What are the main ways that automated decision-making is used at EU borders? Does such automated decision-making bring risks related to human rights, and if so: which risks? The paper introduces a taxonomy of four types of ADM systems at EU borders. Three types are used at borders: systems for (1) identification and verification by checking biometrics, (2) risk assessment, and (3) border monitoring. In addition, (4) polygraphs and emotion detectors are being tested at EU borders. We discuss three categories of risks of such automated decision-making, namely risks related to the human rights to (1) privacy and data protection, (2) non-discrimination, and (3) a fair trial and effective remedies. The paper is largely based on a literature review that we conducted about the use of automated decision-making at borders. The paper combines insights from several disciplines, including social sciences, law, computer science, and migration studies.

## 1 Introduction

Many countries use automated decision-making in the context of international migration and border management. In short, we speak of automated decision-making (ADM), when computers make decisions about people. ADM can include systems driven by artificial intelligence (AI), but ADM systems can also use simpler IT systems. ADM can revolutionize the way that states and international organisations manage international migration and border passing. For instance, ADM can automate human tasks such as identity checks, border surveillance, and processing of visa and asylum applicants' data. But a growing body of scholarship is raising concerns about the use of ADM in international migration and border management (e.g. Broeders and Hampshire 2013, Nalbandian 2022, Nedelcu and Soysüren 2022, Martins and Jumbert 2022, Scheel 2024).

This paper explores the following research questions. (i) What are the main ways in which automated decision-making is or can be used at EU borders? (ii) Does such automated decision-making bring risks related to human rights, and if so: which risks?

Geographically, we focus on Europe. However, given similar technological developments in other regions of the world, the paper is also relevant for readers outside Europe. The term EU borders refers to internal and external borders of the EU. Some topics are outside the scope of the paper. For example, we refrain from critiquing EU border management in general. We only focus on the use of ADM in the EU border management context, and we allude to potential risks for human rights.

The paper makes the following contributions to the literature. First, this paper, based on a multi-disciplinary literature review, integrates insights from various disciplines, including social science, law, computer science, and migration studies. Second, we introduce a taxonomy of four purposes for which ADM is applied in the context of migration and border management: (1) identification and verification by checking biometrics; (2) risk assessment, and (3) border monitoring. (4) Polygraphs and similar tools for lie and emotion detection are being tested at EU borders. Third, we discuss the three most prominent human rights related risks that these four types of ADM incur: risks related to the right to (1) privacy and data protection, (2) non-discrimination, and (3) fair trial and effective remedies. We also highlight provisions in human rights treaties and EU law that are relevant to those risks.

For the literature review, we searched academic literature in electronic databases through the library of a research university in the Netherlands. We searched databases for the phrase 'automated decision-making' but also for related phrases 'artificial intelligence', 'algorithmic decision-making', 'digital borders', and 'data', in combination with the phrase 'border control'. We enriched the literature collection with grey literature, such as research reports. We also asked for literature tips from experts in the field. We workshopped the paper at several conferences and other academic events; the discussions there helped us to improve the paper and to find more relevant literature.

The paper is structured as follows. Section 2 introduces the large-scale IT systems with data on travellers, developed for different purposes, including ADM at the borders of Europe. Section 3 then provides a taxonomy of four categories of ADM at borders.

Section 4 discusses three categories of risks for human rights. In section 5 we reflect on the findings and provide suggestions for further research. Section 6 concludes by answering the research questions.

## 2    Large scale IT systems with data on non-EU nationals

The EU hosts a variety of large-scale IT systems with data on non-EU nationals, each tailored to support specific domains and functions crucial to the EU's operations. The IT systems work together to support the EU's overarching goals of enhancing security, facilitating cooperation, and promoting the free movement of people across member states. The number of databases, IT systems, and ADM systems for EU border management is overwhelming, and their workings are complicated. Below we highlight the main systems.

Three systems have been in place for years: the *Schengen Information System (SIS)*, the *European Asylum Dactyloscopy Database (Eurodac)*, and the *Visa Information System (VIS)*. In addition, the *Entry/Exit System (EES)* and the *European Travel Information and Authorisation System (ETIAS)* are under development. Furthermore, the *European Criminal Records Information System for Third Country Nationals (ECRIS-TCN)* will be integrated into these IT systems to form the building blocks of the new EU Interoperability Initiative (Regulation EU 2019/817 and 2019/818; Brouwer 2020). EU large-scale IT information systems aim at collecting, processing, and storing biometric data (except ETIAS) and biographic data (except Eurodac) of a large number of third-country nationals (Ferraris 2022).

Originally, these IT systems were built for narrower purposes. Over the years and via various legislative modifications, the EU expended the number of purposes associated with each system. The systems have steadily become more focused on border control, granting law enforcement agencies greater access for security purposes. The main purpose of the six main IT systems (planned to be) used at EU borders can be seen in Table 1.

| IT System | Data Type | Purpose | Operational period |
|---|---|---|---|
| **SIS** | Biometric data (fingerprints, facial images, palm prints, and, only for the purpose of police and judicial cooperation, DNA data) and biographical data (e.g. data on lost or stolen objects) | Storing alerts on alerts on third-country nationals subject to a refusal of entry and stay, alerts on persons wanted for arrest for surrender purposes or extradition purposes, alerts on missing persons, alerts on persons sought to assist with a judicial procedure and alerts on persons for discreet checks or specific checks. With the aim of ensuring a high level of security within the area of freedom, justice, and security recording data for the purpose of refusal of entry, the facilitation of return of irregular migrants, and police and judicial cooperation in criminal matters. | In operation since 1995; SIS II in operation since 2013. A renewed SIS with new alerts, upgraded data and enhanced functionalities was launched in March 2023. |
| **Eurodac** | Biometric data (fingerprints, facial images) | Managing asylum applications establishing the responsibility of a Member State for the asylum application; contributing to the fight against irregular | In operation since 2003 and reformed since then. EU lawmakers are discussing a new reform of Eurodac. |

| | | migration, and since 2013, law enforcement purposes. | |
|---|---|---|---|
| **VIS** | Biometric data (fingerprints and facial images) and visa-related data | The exchange of data on short-stay visas between Member States facilitating the visa procedure, preventing visa fraud and irregular immigration, and preventing threats to the internal security of the Member States. | In operation since 2011. The EU adopted a new regulation in 2021 allowing the interoperability with and querying of other EU large-scale databases and Interpol and allowing for the use of algorithm-based risk indicators and the inclusion of data of minors over the age of 6. |
| **ECRIS-TCN** | Biometrics (fingerprints and in some cases facial images) and previous criminal records | Facilitating the exchange of criminal records on convictions of third country nationals, stateless persons, and EU nationals with double nationality (EU and non-EU nationality) between judges and prosecutors within the EU to identify the Member State holding criminal records information on the registered persons to support their statutory tasks. | Established in 2012 as a decentralised IT system based on national criminal record databases. In 2019 on the law ECRIS-TCN was adopted. The system is under development. |
| **EES** | Biometric data (fingerprints and facial images) and entry and exit data (e.g., date and place) | Recording and storage of date, time and place of entry and exit of third–country nationals crossing the borders of the Member State at which the EES is operated with the objective ofenhancing the efficiency of border checks and the calculation of the duration of authorised stay, combating identity fraud and misuse of travel | The EU adopted the regulation for an entry/exit system in 2017. Planned to be operational in 2024. |
| | | documents, and contributing to the prevention, detection, investigation and prosecution of terrorist offences and serious crime. | |
| **ETIAS** | All data included in the individual ETIAS application forms (incl. name, nationality, travel document data, email address, home address, education, current occupation, Member State of first intended stay and optionally address of first intended stay, including data on minors, and information in response to questions concerning whether the person has been convicted over the last ten years for a criminal offence listed in the annex to the ETIAS Regulation or whether he/she has stayed in a specific war of conflict zone over the last ten years and the reasons for the stay. | Storage of information on visa-exempt non-EU nationals applying for an ETIAS enabling national authorities to establish before their arrival whether presence would pose a security, illegal immigration or pandemic risk and allowing access to designated authorities and Europol for the purposes of the prevention, detection, investigation and prosecution of terrorist offences and serious crime.<br><br>Includes the use of ETIAS screening rules based on algorithmic risk profiling and ETIAS watch list of persons who are suspected of having committed or taken part in a terrorist offence or other serious criminal offence or persons regarding whom there are factual indications or reasonable grounds, based on an overall assessment of the person, to believe that they will commit a terrorist offence or other serious criminal offence. | The EU adopted the regulation for the ETIAS in 2018. Planned to be operational in 2025. |

Table 1. Main IT systems with data on non-EU nationals for EU border control.

In May 2019, the EU adopted two regulations to establish a framework for interoperability between the EU large-scale databases. According to the EU Commission, 'a single, overarching EU information system with multiple purposes would deliver the highest degree of information sharing' (European Commission 2010, 3). Based on this new interoperability system, national authorities and EU agencies may gain access to the biometric information of all non-EU nationals, including asylum applicants, visa holders, undocumented migrants and visa-exempt travellers (Ferraris 2022; Quintel 2020).

The framework for interoperability of EU information systems aims to combat identity fraud and to simplify accurate identification of individuals. The regulations establish four interoperability components, as follows: (1) the European Search Portal (ESP), which would enable authorities to search stored biometric and biographical data in multiple information systems simultaneously. (2) a Biometric Matching Service (BMS), which would search and cross-check biometric data from multiple systems to detect potential identity fraud. (3) a Common Identity Repository (CIR), which would store the biographical identity data of non-EU citizens available in multiple EU information systems. (4) a Multiple Identity Detector (MID), which would detect multiple identities linked to the same biometric data (Brouwer 2020; Vavoula 2020). This new interoperability architecture aims for higher efficiency by providing fast, seamless and systematic access to the information of cross-borders to end-users (e.g., border guards, immigration officials). In conclusion, there are various databases and large-scale IT systems with data on non- EU nationals.

### 3 Four categories of automated decision-making at EU borders

This section summarizes four categories of automated decision-making, all of which use the above-mentioned databases and IT systems. As noted, we speak of automated decision-making, or ADM, when computers make decisions about people. Hence, ADM refers to the use of computers, data, and algorithms to make decisions. ADM can rely on reasonably simple software, or on sophisticated AI. AI can be described as 'the science and engineering of making intelligent machines, especially intelligent computer programs' (McCarthy 2007, 2).

Sometimes, a computer can decide about a person fully automatically, for instance when an e-gate (and automated border crossing gate) checks a traveller's passport and allows them to pass. Sometimes, a human, for instance an immigration officer, receives advice, or a suggested decision, from a computer. For example, a computer may flag a traveller as suspicious, and an immigration officer decides, partly based on that digital flag, whether the traveller can pass. In this paper, we use the phrase ADM for both fully and partly automated decision-making. Below, we discuss four categories of ADM systems, namely systems for (1) identification and verification by checking biometrics, (2) risk assessment, (3) border monitoring and (4) polygraphs or emotion detectors.

### 3.1 Biometric systems and e-gates for identification and verification

A first category of ADM systems checks, automatically, a travellers' biometric data, such as fingerprints for the purpose of identification and/or verification. E-gates, also called automated border control systems, are used extensively. E-gates are a type of automated self-service barriers. E-gates using face recognition are used at EU airports, including those in Italy, Portugal, and The Netherlands (European Parliament. Directorate General for Parliamentary Research Services and Dumbrava 2021).

E-gates usually combine data that they gather with data they read from a chip in biometric passports to verify the identity of the travellers (i.e., biometric verification). E-gates often query border control records stored in databases to determine the eligibility of border crossing based on the pre-defined rules (e.g., biometric identification, the effective date of e-passport) (European Parliament. Directorate General for Parliamentary Research Services and Dumbrava 2021; Petrou, Ntantogian, and Xenakis 2013; Sager 2023; Saunders 2023a).

Typically, there are three steps when a traveller passes an e-gate: (1) the system validates the authenticity of the documents (e.g., the e-passport); (2) the system verifies the identity of the travellers using biometrics (such as fingerprints or a picture of the traveller's face); and (3) the system verifies the validity of the traveller's authorisation (e.g., the visa) (Labati et al. 2015). Border officials manually check the traveller's identification if the automated biometric recognition fails.

Matching algorithms are widely used at the borders for biometric identity verification and identification (Saunders 2023a; 2023b). European e-passports store face and fingerprint data, so ADM systems can check whether the traveller is the person whose name is on the e-passport (Labati et al. 2015).

Some e-gates use automated fingerprint recognition. However, there are still some challenges for fingerprint-sensing solutions that constrain the applicability of this biometric technology. For instance, fingerprint systems might work badly for children, or for people who have worn-out fingerprints such as some guitarists (Sousedik et al. 2017).

ADM systems sometimes use other biometric traits, such as iris recognition. For some e-gates, using an iris recognition system is optional, for instance in Amsterdam (Labati et al. 2015). ADM systems could also recognise a person's body shape. As far as we know, such systems are not in use at EU borders (Vera-Rodriguez et al. 2017). In sum, various types of e-gates and biometric systems for identification and verification are being used at European borders.

### 3.2 Risk assessment

A second type of ADM system performs automated risk assessments of travellers. To check whether a traveller is authorized to cross the border, the ADM system checks their data for risks, such as identity fraud, criminality or terrorism (La Fors and Meissner 2022; Saunders 2023a). In Europe, airlines provide Passenger Name Record (PNR) data, which are analysed by ADM systems. (The United States uses a similar system: the Advanced Passengers Information System.)

For flights from outside the Schengen area, airlines collect both PNR data (generated at the booking stage) and Advance Passenger Information (API) data (collected by airlines at check-in). The Schengen area is, roughly summarised, a group of 27 countries in Europe that have abolished checks at the internal borders. The EU PNR system provides for the analysis of passenger data from inbound and outbound extra-EU flights, including information on the itinerary, the name of the passenger, ticket details, the name of the person who made the booking, etc. (PNR Directive 2016). All EU Member states except one have extended the collection of PNR data to intra-EU flights (European Commission 2020).

Airlines and other companies transmit the data to the national Passenger Information Unit (PIU) before flight closure and function as a 'pre-border' for automated processing (Sager 2023). The PIU screens all travellers to an EU port of entry, including EU and Schengen state citizens, against profiles or against names contained in law enforcement or security watchlists (Jeandesboz 2021; Saunders 2023b).

For automated risk assessment, as we have seen above, the purpose of existing databases has been extended, and new databases are being developed. The new ETIAS will profile visa-exempt third country nationals using a screening rule algorithm to make automated predictive risk assessments (Musco Eklund 2023). Gradually, a large EU-wide information system is being developed (Ferraris 2022).

Personal data are widely shared between government departments and across states in the EU's integrated border management. Several EU agencies play a role regarding the use and the management of data in EU's large scale IT systems. *Europol* has access to VIS, SIS and EURODAC and when operational, EES, ETIAS and ECRIS-TCN (Bigo 2014; Trauttmansdorff and Felt 2021). Information in the latter system, ECRIS-TCN, is also accessible for Eurojust and the European Public Prosecutor's Office. Frontex, the European Border and Coast Guard Agency, has access to VIS and EES for the purpose of carrying out risk analyses and vulnerability assessments. Furthermore, the ETIAS Central Unit, responsible for the correct use and storage of data in the ETIAS system, is established within *Frontex*. eu-LISA is responsible for the operational management of EU's large-scale IT systems, including their interoperability (Valdivia et al. 2022). Frontex is exploring the development of information systems that interconnect the national police and coast guard authorities (see more details in section 3.3, Pollozek and Passoth 2023). In addition, eu-LISA and Frontex formed working agreements and collaboration (formally established in 2014) to enhance joint analysis (Valdivia et al. 2022). Other agencies, such as Europol, have garnered international cooperation beyond the EU (Coman-Kund 2020). In sum, the EU is aiming to combine multiple IT systems for risk assessment at its borders.

### 3.3 Border monitoring

A third category of automated systems is used for border surveillance. The European Border and Coast Guard Agency, commonly referred to as Frontex, uses systems such as video surveillance from planes and drones (Pollozek 2020). Besides, other vision-based systems, such as a multi-sensor fusion and tracking system and unmanned aerial vehicles, are tested to automatically detect irregular border-crossers (Avola et al. 2019; Patino et al. 2022). The EU sometimes gives research funding to develop surveillance technology (e.g., the Horizon 2020 FOLDOUT project) (Patino et al. 2022).

Frontex assists EU Member States and Schengen-associated countries in the EU's external border management and the fight against cross-border crimes. As Frontex puts it, 'our Maritime Aerial Surveillance (MAS) has become an integral part of our operations, and a permanent service offered to national authorities. MAS uses surveillance airplanes and drones that stream video and other data from EU and Schengen external borders directly to our headquarters in Warsaw and to national and European authorities, allowing for real-time monitoring' (Frontex no date). Border monitoring is not fully automated. At the European Monitoring Room in Warsaw, specialists monitor the incoming information. EU member states can request Frontex to operate planes and drones to perform coastal surveillance functions. Frontex aircrafts carried out 1,000 missions in 2022 (Frontex 2022).

The EU has invested substantially in IT systems in the past decade, such as the Joint Operation Reporting Application (JORA) and the European Surveillance System (Eurosur) (Pollozek 2020). The Joint Operation Reporting Application (JORA) is an information system of Frontex, in use since 2011 in the Mediterranean. JORA is designed for border monitoring and migrant identification and registration (Leese and Pollozek 2023). Eurosur, established in 2013, 'is a framework for information exchange and cooperation between Member States and Frontex to improve situational awareness and increase reaction capability at the external borders' (European Commission no date). Eurosur links various information systems at the local, national, and European level. The Eurosur Fusion Services include automated vessel tracking and detection capabilities, automated anomaly identification and vessel location prediction (Bellanova and Duez 2016; Frontex no date). In conclusion, various fully and partly automated systems are used to monitor EU borders.

### 3.4 Polygraphs and emotion-detectors

A fourth category of ADM systems aims to detect emotions or intention by using AI-based assessments of people's physical features. Presently there are no polygraphs (lie detectors) or emotion-detectors deployed at EU borders. But several EU-(co)funded projects, including Automated Virtual Agent for Truth Assessments in Real-Time (AVATAR) and Intelligent Portable Border Control System (iBorderCtrl), have piloted such technologies with the objective of detecting the intentions and emotions of individuals throughout the migration process.

Emotion detection systems aim to identify somebody's emotional state, intention, or mental state, such as deception or trustworthiness. Emotion detection technology uses a person's biometric data, or their physiological, physical, or behavioural characteristics, such as gaze direction, voice, gesture, heartbeat rate, body or face temperature, and skin conductivity (European Parliament. Directorate General for Parliamentary Research Services and Dumbrava 2021).

For example, the iBorderCtrl pilot project included, in short, a polygraph. The research project wrote that 'iBorderCtrl provides a unified solution with aim to speed up the border crossing at the EU external borders and at the same time enhance the security and confidence regarding border control checks by bringing together many state of the art technologies (hardware and software) ranging from biometric verification, *automated deception detection*, document authentication and risk assessment' (italics by the authors) (iBorderCtrl 2016). The system was created with financial support from the European Commission's Horizon 2020 program to help with border control. The aim of the iBorderCtrl pilot was to establish whether a traveller lies during an immigration interview, by measuring micro-expressions that supposedly show when a person is lying. The iBorderCtrl system uses 38 features, such as 'left eye blink' and 'increase in facial redness,' to detect deception (Sánchez-Monedero and Dencik 2022, 418). Pilot testing has been conducted in Hungarian, Greek, and Latvian land borders (Sánchez-Monedero and Dencik, 2022). In section 4.3 we discuss criticism on emotion and lie detection systems.

In conclusion, many ADM systems are used or tested at EU borders. Three categories of ADM systems that are in use are (1) systems for automated biometric identification or verification, such as e-gates, (2) risk assessment systems, and (3) border monitoring tools. The EU funds research into (4) automated lie and emotion detectors for borders. The next section explores whether such automated decision-making at borders bring risks for human rights, and if so: which risks?

## 4  Risks for human rights of automated decision-making at EU borders

This section discusses risks of ADM systems at borders. We discuss three broad categories of risks for human rights of travellers, namely risks related to the right to (1) privacy and data protection, (2) non-discrimination, and (3) a fair trial and effective remedies. We highlight international human rights standards that relate to the risks.

Table 2. gives a non-exhaustive overview of the main risks related to human rights of the four applications of ADM systems.

|  | **Privacy and data protection** | **Discrimination** | **Fair trial/effective remedies** |
|---|---|---|---|
| **Biometric systems** | Use of sensitive data; long term data retention; repurposing of personal data. | May work badly for certain ethnicities or other groups. | Reasons for decision can remain secret. |
| **Risk assessment** | Repurposing of personal data; lack of transparency; incentive for large-scale data collection; stigmatizing effect. | Profiling based on e.g. ethnicity; application of ADM mostly on individuals belonging to certain nationalities, ethnicities, or backgrounds. | Reasons for decision can remain secret, hampering accessibility and effectiveness of legal remedies. |
| **Border monitoring tools** | Repurposing of personal data; lack of transparency; incentive for large-scale data collection; stigmatizing effect of border surveillance; data sharing with private sector or third countries. | Application of ADM mostly on individuals belonging to certain nationalities, ethnicities, or backgrounds. | Reasons for decision can remain secret, hampering accessibility and effectiveness of legal remedies. |
| **Polygraphs or Emotion detector** | Use of sensitive data; unreliability; stigmatizing effect. | May work badly for certain ethnicities or other groups; application of ADM mostly on individuals belonging to certain nationalities, ethnicities, or backgrounds. | Unreliability; reasons for decision can remain secret, hampering accessibility and effectiveness of legal remedies. |

Table 2.  The main risks related to human rights of four applications of ADM systems.

## 4.1  Privacy and data protection

The first category of risks relates to privacy and data protection. The right to privacy is included in many human treaties, including the UN International Covenant on Civil and Political Rights (1965, Article 17) and the European Convention on Human Rights (1950, Article 8). Data protection law regulates the processing of personal data. In the European Union, the right to the protection of personal data is also protected as a fundamental right, namely in the Charter of Fundamental Rights of the European Union (2000, Article 8). Article 8 of the Charter also requires that personal data are processed 'fairly for specified purposes and on the basis of the consent of the person concerned or some other legitimate basis laid down by law', which implies transparency about how and for which purposes personal data are used (GDPR 2016, Article 5(1)(a)). (The phrases human rights and fundamental rights can be regarded as synonyms.[1])

We highlight some specific privacy and data protection risks of ADM. Biometric systems (the first category of ADM systems) use and analyse personal data such as fingerprints, pictures of faces, and iris scans. Such data are sensitive, and can disclose, for instance, people's ethnicity or a disease. The use of ADM systems for identification and verification relies upon the use of databases with sensitive data. The mere storage of personal data interferes with privacy and data protection rights as protected in the European Convention on Human Rights and EU Charter of Fundamental Rights (Court of Justice of the European Union 2014, 2022; European Court of Human Rights 2021). Moreover, collecting and storing of biometric data of children in Eurodac and VIS impacts rights and interests of children as protected in the Convention on the Rights of Children and the EU Charter of Fundamental Rights.

Second, ADM systems for risk assessments also interfere with privacy and data protection rights. For automated risk assessments, large amounts of personal data are collected, stored, combined, and analysed. While all ADM systems incentivize data collection, this incentive is especially large for automated risk assessment systems. At EU borders data are collected from multiple sources, including official government biometrics, private company records and, in some cases, from individuals' smartphones (e.g., in Germany, Austria, Switzerland and the Netherlands) (Josipovic 2024; Sager 2023). If authorities used automated risk assessment tools, they have an incentive to collect and examine increasing amounts of personal data about migrants, including sometimes social media postings, financial transactions, and location data (Leurs 2023; Sager 2023).

---

[1] For the slight difference, see González Fuster 2014, p. 82-84.

Third, border monitoring tools such as drones collect and analyse much personal data, such as video images. The data may also flow to the private sector, for instance through contracts that approve access to collected data (Saunders 2023a; Turculet 2023). This brings extra risks, as there is less democratic oversight on the private sector than on the public sector. To illustrate, the European Data Protection Supervisor (2022) reprimanded Frontex for its migration to cloud computing without conducting a data protection impact assessment to assess the risks.

Fourth, emotion detection and lie detection systems also bring risks for privacy and the protection of personal data. The systems aim to analyse sensitive personal data, including individuals' biometric data, and their physiological, physical, or behavioural characteristics during interviews. It is doubtful whether automated emotion detection could work at all (see section 4.3). Assuming that it could work, it would have serious privacy implications if an automated system could read people's emotions.

More generally, most ADM systems rely on using large amounts of personal data. Therefore, the four categories of ADM systems at borders all incentivize data collection and storage, and thus bring privacy risks (Leurs 2023; Perret and Aradau 2023).

## 4.2 Non-discrimination

The second category of risks of ADM concerns discrimination and inequality. The right to non-discrimination is protected, for instance, in the UN International Covenant on Civil and Political Rights (1965, Article 26), the European Convention on Human Rights (1950, Article 14), and the Charter of Fundamental Rights of the European Union (2000, Article 21).

First, biometric systems bring discrimination risks. For example, some facial recognition systems perform worse for certain ethnicities. To illustrate, for a dataset of high-quality photos (for visa applications), 'false positive rates are highest in West and East African and East Asian people, and lowest in Eastern European individuals' (Grother, Ngan, and Hanaoka 2019, 2). In this context, '(f)alse positives are the erroneous association of samples of two persons; they occur when the digitized faces of two people are similar'. False positives also happen more often for women, and for certain age groups (Grother, Ngan, and Hanaoka 2019). Systems that should recognize fingerprints can also perform badly for certain groups, such as children (see section 3.1).

If facial recognition technology works badly for dark-skinned people, it seems plausible that ADM systems that aim for emotion or lie detection make more errors for such groups too. We show in the next section that it is generally doubtful whether automated systems can recognise emotions or lies.

Second, automated risks assessment also leads to discrimination risks. For instance, in the Netherlands, the foreign affairs ministry uses a profiling system since 2015 that uses variables like nationality, gender, and age, to compute the risk score of millions of short-stay visa applicants applying to enter the Netherlands and Schengen area. The system flags some applicants with a 'high risk' score. Dutch authorities investigate those applicants, which can lead to months of delay for applicants. For example, it is difficult for family members of Dutch citizens from Morocco and Suriname to obtain a visa (Maleeyakul et al. 2023).

If ADM systems are trained on historical data (such as non-automated decisions by immigration officers, for example in visa procedures) to recognize potential irregular immigrants such as visa overstayers, there is a risk that the ADM system reproduces discrimination or selection underlying those human decisions (Derave, Genicot, and Hetmanska 2022). It is a well-known problem that AI and ADM systems can reproduce or reinforce discrimination in society. Biased training data is one possible cause. Research shows that immigration officers sometimes use ethnic profiling in their border control procedures (Himanen 2022). If such data are used to train ADM systems, the ADM systems may also become discriminatory. In sum, there is a risk of discriminatory effects when ADM is used for risk assessment (Beduschi 2020; Sager 2023).

Third, in a broader sense, ADM systems can be applied in a discriminatory way. ADM systems are often applied only to certain groups, such as people from outside the EU.

For instance, border officers often require vulnerable travellers, including asylum seekers, to hand in their mobile phones to collect their personal data including social media postings, financial transactions, and location to complete risk assessments. Asylum seekers are also sometimes subjected to the experimentation of various high-risk technologies, such as emotion detection (Leurs 2023; Sánchez-Monedero and Dencik 2022). Such differentiation between different travellers can make vulnerable groups the 'testing ground' for high-risk technologies (Molnar 2020; Ozkul 2023). To sum up, the use of ADM systems at borders brings various discrimination-related risks.

### 4.3 The right to fair trial and effective remedies

The third category of risks concerns the risk that automated procedures for immigrants and travellers affect their right to effective judicial protection, due to opaque decision-making and the lack of transparency. The right to a fair trial and effective remedies is protected in international conventions and EU law. For instance, the European Convention on Human Rights (1950) protects the right to a fair trial and the right to an effective remedy (Article 6 and 13). The Charter of Fundamental Rights of the European Union (2000) includes the right to effective judicial protection in Article 47.

ADM systems at borders bring the risk of opaque or otherwise unfair procedures. For instance, when a traveller is subjected to automated risk assessment, the ADM system may flag the person as a risk. After such an automated flag, immigration officers may subject the traveller to more extensive inspections or interrogations. It can be unclear for travellers why the ADM system flagged them as risky.

A well-known problem with ADM and AI systems is their 'black box' character (Amilevicius 2021). Sometimes it is not even clear for developers why an ADM system arrives at certain decisions or outcomes (Brouwer 2021; Saunders 2023a). And even if the developers and users of ADM systems know what is behind a decision, that may not provide transparency to the traveller. Moreover, automated risk assessment systems can make mistakes (Møhl 2022). For example, an innocent tourist could be flagged as a potential terrorist. In 2022, the Court of Justice of the European Union, banned, roughly summarised, fully automated risk assessment based on PNR data ('artificial intelligence technology in self-learning systems'), because of the risks for human rights (Court of Justice of the European Union 2022, para. 194; Gerards 2023; Haitsma 2023).

Emotion detection and lie detection systems include further risks for the accessibility and effectiveness of legal remedies. Automated emotion and lie detection are among the most controversial applications of ADM and AI. Many scholars question whether it is possible at all to detect emotions automatically (Stark and Hutson 2021). Indeed, Barrett et al (2019, 46) conclude in a meta study of scholarly work about detecting emotion from facial expressions: 'It is not possible to confidently infer happiness from a smile, anger from a scowl, or sadness from a frown, as much of current technology tries to do when applying what are mistakenly believed to be the scientific facts.'

The idea that computers can detect whether people lie is perhaps even more controversial. Developers of automated polygraphs or lie detectors appear to assume that there are universal ways of expressing deception through nonverbal cues that apply to all people regardless of ethnicity, age, gender, or neurodiversity. However, this assumption is questionable. As Sánchez-Monedero and Dencik (2022, 416) note,

'Deception or lie detection devices, mainly represented by the polygraph, have always been contentious and lacking in substantial scientific evidence'. It is unfair if somebody is accused by an ADM system of lying, while the system is unreliable.

Indeed, many have argued that the EU should have banned AI-driven polygraphs and emotion detectors in the AI Act, adopted in the spring of 2024 (e.g. Meijers Committee 2022). But the EU lawmakers did not ban such systems. The lawmakers did, however, categorise such systems as 'high risk'; therefore, reasonably strict rules apply to such systems (EU AI Act 2024).

In conclusion, ADM systems at borders threatens human rights. The risks can be categorised in three broad categories: risks related to (1) privacy and data protection (2) non-discrimination, and (3) a fair trial and effective remedies.

## 5    Discussion and suggestions for further research

In this section we reflect on the main findings of this study and provide some suggestions for further research. In the previous section, we mentioned three categories of risks related to ADM at borders. That overview of risks is not exhaustive. Future research could arrive at a more detailed and more extensive analysis of the risks. For instance, we did not analyse the 'militarisation of European borders' or the influence of the private sector on ADM at EU borders (Jones and Johnson 2016).

Because of length constraints, our paper did not assess whether specific ADM systems comply with all the legal requirements, including criminal law procedural rights, best interest of the child, or more specific data protection law standards. Such an assessment would require a detailed analysis of each system separately. Such in-depth legal analysis of the different ADM systems would be interesting, and necessary to prevent risks for human rights.

There are many possibilities for more empirical research to provide further understanding on how the various ADM systems are used. For instance, how effective are different ADM systems at borders? And how much money does the EU and individual member states spend on ADM at borders? Perhaps researchers could obtain previously undisclosed documentation about the ADM systems by doing 'freedom of information' requests to the EU or EU Member States. In addition, the increasing incorporation of ADM and AI into surveillance technologies at EU borders prompts inquiries into its effects on the human experiences. How do people experience the surveillance and ADM systems at borders?

As noted, the EU funds controversial projects such as iBorderCtrl. Meanwhile, that the EU has also funded research that is critical of iBorderCtrl and similar systems (e.g. Sánchez-Monedero and Dencik 2022). There is some ethnographic work on certain ADM systems at EU borders (e.g. Møhl 2022; Simonsen 2022). More ethnographic work would be welcome. This type of research can provide insights into the use and effects of ADM at borders (Martins and Jumbert 2022).

Although AI and other ADM systems can enhance border security, improve efficiency, and streamline operations in many ways, such systems also bring risks for human rights, as illustrated in this paper. The use of ADM at borders can harm vulnerable groups and reinforce social inequality. Such effects can also happen in other contexts than borders. For example, many states experiment with automated systems for fraud detection in the welfare state (Ozkul 2023). There is a risk that such systems harm poor or otherwise vulnerable groups in society. ADM can also reinforce societal inequality in other ways, for example by strengthening the market power of large tech companies. More research on questions related to ADM and inequality is necessary to understand the harms to vulnerable groups, also outside the immigration context. Perhaps debates about ADM in the immigration context could learn from debates in other contexts, and vice versa.

## 6   Conclusion

We conclude by answering the main research questions. We explored the main ways in which automated decision-making systems are used at EU borders, and the risks of such systems. We introduced a taxonomy of four types of ADM systems at borders. Three types are used at EU borders: systems for (1) identification and verification by checking biometrics, (2) risk assessment, (3) and border monitoring. In addition, (4) polygraphs and emotion-detectors are being tested at EU borders. The four ADM types for migration and border control raise multiple social, ethical, and legal concerns. We distinguished three broad categories of human rights risks of such ADM, namely risks related to (1) privacy and data protection, (2) non-discrimination, and (3) a fair trial and effective remedies. We call for more research in various disciplines, more public debate, and more democratic oversight of automated decision-making at EU borders.

* * *